
%
%
\input phyzzx
\def\etal{{\it et al.}}

\def\twog{\gamma\gamma}
\def\threeg{\gamma\gamma\gamma}
\def\ll{\ell^+\ell^-}
\def\to{\rightarrow}
\def\sigmaX{\sigma_{\lower 2pt\hbox{$\scriptstyle X$}}}
\def\sigmaE{\sigma_{\lower 2pt\hbox{$\scriptstyle E$}}}
\def\Gee{\Gamma_{\rm ee}}
\def\Ggg{\Gamma_{\twog}}
\def\BRgg{{\rm BR}_{\twog}}
\def\keV{\,{\rm keV}}
\def\MeV{\,{\rm MeV}}
\def\GeV{\,{\rm GeV}}

\def\ee{{\rm e^+e^-}}

\def\Z0{\rm Z^{\circ}}

\hsize=6in
\hsize=6.0in
%
\REF{\rfLTHREE}{O.~Adriani \etal\  (L3 Collaboration), Phys.~Lett.~{\bf 295B}
(1992) 337.}
\REF{\rfLEP}{S.~Komamiya, K.W.~Bell and S.~Holmgven, private communications.}
%
\REF{\rfTRIUMF}{R.~Garisto and J.N.~Ng, Preprint TRI-PP-92-124, 16
Dec.~1992.}
\REF{\rfKEK}{K.~Hagiwara and S.~Matsumoto, private communications.}
%
\REF{\rfVENUS}{K.~Abe \etal, Z.~Phys.~C {\bf 45} (1989) 175.}
\REF{\rfTOPAZ}{K.~Shimozawa \etal, Phys.~Lett.~{\bf B284} (1992) 144.}
\REF{\rfAMY}{AMY Collaboration, T.~Kumita \etal, Phys.~Rev.~D24 (1990) 1339.}
\REF{\rfSHC}{A.~Abashian \etal, to be published in Nucl.~Instr.~and Meth.~A.}
\REF{\rfFUIG}{J.~Fujimoto and M.~Igarashi, Prog.~Theor.~Phys.~{\bf 74} (1985)
791.}
\REF{\rfFUIGSH}{J.~Fujimoto, M.~Igarashi, Y.~Shimizu, Prog.~Theor.~Phys.
 ~{\bf 77} (1987) 118.}
\REF{\rfSKAW}{S.~Kawabata, Comp.~Phys.~Comm.~{\bf 41} (1986) 127.}
\REF{\rfJACKSON}{J.D.~Jackson and D.L.~Scharre, Nucl.~Instr.~and
 Meth.~{\bf 128} (1975) 13.}
\REF{\rfHELENE}{O.~Helene, Nucl.~Instr.~and Meth.~{\bf 212} (1983) 319.}
\REF{\rfKURIHARA}{Y.~Kurihara (unpublished).}
\REF{\rfGRACE}{H.~Tanaka, Comp.~Phys.~Comm.~{\bf 58} (1990) 153, H.~Tanaka,
T.~Kaneko, and Y.~Shimizu, Comp.~Phys.~Comm.~{\bf 64} (1991) 149, T.~Kaneko,
in ``New Computing Techniques in Physics Research,'' ed.~Perret-Gallix
and W.~Wojcik, (Edition du CRNS, Paris: 1990) 555.}
\REF{\rfOPAL}{M.Z.~Akrawy \etal\ (OPAL Collaboration), Phys.~Lett.~{\bf 257B}
 (1991) 531.}
\REF{\rfAGUIL}{F.~del \'Aguila, A.~M\'endez and R.~Pascual,
 Phys.~Lett.~{\bf 140B} (1984) 431.}
\REF{\rfKINOSH}{T.~Kinoshita and W.B.~Linquist, Phys.~Rev.~Lett.~{\bf 47}
 (1981) 1573.}
%
\vskip 0.2in
\centerline{\bf Search for Anomalous $\twog$ Production at TRISTAN}
\vskip 0.2in
%
%
%
\singlespace \twelvepoint
\newbox\smallstrutbox
\setbox\smallstrutbox=\hbox{\vrule height10.5pt depth3.5pt width0pt}
\def\smallstrut{\relax\ifmmode\copy\smallstrutbox\else\unhcopy
  \smallstrutbox\fi}
{\tenrm\offinterlineskip}
\singlespace \tenpoint
\centerline{\bf The AMY Collaboration}
\tolerance=10000
\def\vpi{a}	
\def\lsu{b}	
\def\beij{c}	
\def\usc{d}	
\def\ssc{e}     
\def\ucd{f}	
\def\uh{g}      
\def\kek{h}	
\def\konan{i}	
\def\mines{j}	
\def\niig{k}	
\def\niden{l}	
\def\rutg{m}	
\def\ur{n}	
\def\saga{o}	
\def\kore{p}	
\def\seoul{q}	
\def\knu{r}	
\def\chuo{s}	
\def\sait{t}	
\noindent
%
K.L.~Sterner,\attach{\rm \vpi}
A.~Abashian,\attach{\rm \vpi}
K.~Gotow,\attach{\rm \vpi}
D.~Haim,\attach{\rm \vpi}
M.E.~Mattson,\attach{\rm \vpi}
N.~Morgan,\attach{\rm \vpi}
L.~Piilonen,\attach{\rm \vpi}
%
P.~Kirk,\attach{\rm \lsu}
%
C.P.~Cheng,\attach{\rm \beij}
W.X.~Gao,\attach{\rm \beij}
W.G~Yan,\attach{\rm \beij}
M.H.~Ye,\attach{\rm \beij}
%
S.~Lusin,\attach{\rm \usc}
C.~Rosenfeld,\attach{\rm \usc}
A.T.M.~Wang,\attach{\rm \usc}
S.~Wilson,\attach{\rm \usc}
L.Y.~Zheng,\attach{\rm \usc}
%
%
C.A.~Fry,\attach{\rm \ssc}
R.~Tanaka,\attach{\rm \ssc}      
R.E.~Breedon,\attach{\rm \ucd ,\kek}
L.M.~Chinitz,\attach{\rm \ucd}
Winston~Ko,\attach{\rm \ucd}
R.L.~Lander,\attach{\rm \ucd}
J.~Rowe,\attach{\rm \ucd}
J.R.~Smith,\attach{\rm \ucd}
D.~Stuart,\attach{\rm \ucd}
%
S.~Kanda,\attach{\rm \uh}
S.L.~Olsen,\attach{\rm \uh}
K.~Ueno,\attach{\rm \uh}
K.~Abe,\attach{\rm \kek}
Y.~Fujii,\attach{\rm \kek}
Y.~Kurihara,\attach{\rm \kek}
F.~Liu,\attach{\rm \kek}
A.~Maki,\attach{\rm \kek}
T.~Nozaki, \attach{\rm \kek}
T.~Omori,\attach{\rm \kek}
H.~Sagawa,\attach{\rm \kek}
Y.~Sakai,\attach{\rm \kek}
T.~Sasaki,\attach{\rm \kek}
Y.~Sugimoto,\attach{\rm \kek}
Y.~Takaiwa,\attach{\rm \kek}
S.~Terada,\attach{\rm \kek}
R.~Walker,\attach{\rm \kek ,\ur}
F.~Kajino,\attach{\rm \konan}
%
R.~Poling,\attach{\rm \mines}
T.~Thomas,\attach{\rm \mines}
T.~Aso,\attach{\rm \niig}
K.~Miyano,\attach{\rm \niig}
H.~Miyata,\attach{\rm \niig}
M.~Oyoshi,\attach{\rm \niig}
%
M.H.~Lee,\attach{\rm \rutg}      
F.~Sannes,\attach{\rm \rutg}
S.~Schnetzer,\attach{\rm \rutg}
R.~Stone,\attach{\rm \rutg}
J.~Vinson,\attach{\rm \rutg}
%
Y.~Yamashita,\attach{\rm \niden}
%
A.~Bodek,\attach{\rm \ur}
B.J.~Kim,\attach{\rm \ur}
T.~Kumita,\attach{\rm \ur}
Y.K.~Li,\attach{\rm \ur}
C.~Velisarris,\attach{\rm \ur}
%
S.~Kobayashi,\attach{\rm \saga}
A.~Murakami,\attach{\rm \saga}
S.K.~Sahu,\attach{\rm \saga}
%
Y.S.~Chung,\attach{\rm \kore}
J.S.~Kang,\attach{\rm \kore}
K.B.~Lee,\attach{\rm \kore}
K.W.~Park,\attach{\rm \kore}
S.K.~Kim,\attach{\rm \seoul}
S.S.~Myung,\attach{\rm \seoul}
S.K.~Choi,\attach{\rm \knu}
D.~Son,\attach{\rm \knu}
S.~Ebara,\attach{\rm \chuo}
S.~Matsumoto,\attach{\rm \chuo}
N.~Takashimizu,\attach{\rm \chuo}
%
%
 and
T.~Ishizuka\attach{\rm \sait}
%
\vskip .10in
\newbox\smallstrutbox
\setbox\smallstrutbox=\hbox{\vrule height10.5pt depth3.5pt width0pt}
\def\smallstrut{\relax\ifmmode\copy\smallstrutbox\else\unhcopy
  \smallstrutbox\fi}
{\tensl\offinterlineskip
\centerline{\smallstrut
\attach{\rm \vpi}Virginia Polytechnic Institute and State University,
	Blacksburg, VA 24061, USA}
\centerline{\smallstrut
\attach{\rm \lsu}Louisiana State University, Baton Rouge, LA 70803, USA}
\centerline{\smallstrut
\attach{\rm \beij}Institute of High Energy Physics, Beijing 100039, China}
\centerline{\smallstrut
\attach{\rm \usc}University of South Carolina, Columbia, SC 29208, USA}
%
%
%
\centerline{\smallstrut
\attach{\rm \ssc}SSC Laboratory, Dallas, TX 75237, USA}
\centerline{\smallstrut
\attach{\rm \ucd}University of California, Davis, CA 95616, USA}
\centerline{\smallstrut
\attach{\rm \uh}University of Hawaii, Honolulu, HI 96822, USA}
\centerline{\smallstrut
\attach{\rm \kek}KEK, National Laboratory for High Energy Physics,
 Ibaraki 305, Japan}
%
%
\centerline{\smallstrut
\attach{\rm \konan}Konan University, Kobe 658, Japan}
\centerline{\smallstrut
\attach{\rm \mines}University of Minnesota, Minneapolis, MN 55455, USA}
\centerline{\smallstrut
\attach{\rm \niig}Niigata University, Niigata 950-21, Japan}
\centerline{\smallstrut
\attach{\rm \niden}Nihon Dental College, Niigata 951, Japan}
\centerline{\smallstrut
\attach{\rm \rutg}Rutgers University, Piscataway, NJ 08854, USA}
%
%
\centerline{\smallstrut
\attach{\rm \ur}University of Rochester, Rochester, NY 14627, USA}
\centerline{\smallstrut
\attach{\rm \saga}Saga University, Saga 840, Japan}
\centerline{\smallstrut
\attach{\rm \kore}Korea University, Seoul 136-701, South Korea}
\centerline{\smallstrut
\attach{\rm \seoul}Seoul National University, Seoul 151-742, South Korea}
\centerline{\smallstrut
\attach{\rm \knu}Kyungpook National University, Taegu 702-701,
South Korea}
\centerline{\smallstrut
\attach{\rm \chuo}Chuo University, Tokyo 112, Japan}
%
\centerline{\smallstrut
\attach{\rm \sait}Saitama University, Urawa 338, Japan}
%
\Twelvepoint
\endpage
\centerline{\bf Abstract}
\singlespace

\noindent
We report on measurements of the total cross section for $\ee \to \twog$
for center-of-mass energies between 57.4 and 59.5 GeV, using the AMY
detector at the TRISTAN collider.  We set new limits on the production of
a possible new $s$-channel resonance decaying into photon pairs.
\vskip 1.0in
\rm
\doublespace
\Twelvepoint

\chapter{\bf Introduction}

Recently, the L3 Collaboration at LEP reported on an apparent excess of
$\ee \to \ll\twog$ events ($\ell = \mu\ {\rm or}\ e$) in which the invariant
mass of the photon pair was clustered around $59\GeV$.\refmark{\rfLTHREE}
The L3 Collaboration
did not speculate on the origin of their events beyond noting that the
probability of observing such clustering in any $5\GeV$ wide mass
bin above $40\GeV$ was ${\cal O}(10^{-3})$, if the events were due purely
to QED\null. The three other experiments at LEP have also reported on such
events.\refmark{\rfLEP} Out of fifteen $\ee \to \ll\twog$ events found at LEP
with $M_{\twog} > 40\GeV$, five
events were found with a $\twog$ invariant mass in a 1 GeV range near
59 GeV\null.

A few speculative models have been advanced\refmark{\rfTRIUMF,\rfKEK} on
the origin of the clustered events of L3 assuming that they are not due to
QED\null.  The models are based on the decay of the $Z^{\circ}$ into a massive
object $X$ and another boson, where $X$ may be a scalar-pseudoscalar mixture
or a spin-2 particle with a mass of approximately $60\GeV$ that
decays predominantly into two photons.  Regardless of the details of such
models, if the particle $X$ can couple with electrons, then the direct
$s$-channel production may be observable in the TRISTAN energy range.
If there is no coupling of $X$ to fermions, it can still be produced via
an $s$-channel photon that results in $X \to \twog$ plus a soft monochromatic
photon. Previous studies of $\ee \to \twog,\threeg$ at
TRISTAN\refmark{\rfVENUS,\rfTOPAZ} were consistent with the QED prediction,
however, no direct search for a narrow resonance was conducted.


We have searched for the direct production of a new state $X$ via the
reaction $\ee \to X \to \twog$,
using the
AMY detector at the TRISTAN $\ee$ collider, for center-of-mass energies
$\sqrt{s}$ between 57.4 and 59.5 GeV in
250 MeV steps, with 1 to 2 $\rm pb^{-1}$ of integrated luminosity per point
(19 $\rm pb^{-1}$ at 57.8 GeV)\null.

\chapter{\bf The AMY Detector}

The AMY detector is a general purpose solenoidal-type instrument employing
two inner tracking chambers (VTX and ITC), a central drift chamber (CDC),
and an electromagnetic calorimeter (SHC), all contained within a 3 Tesla
magnetic field, and a barrel muon detector (MUO) outside the magnet return
yoke.  The end cap regions are instrumented with calorimeters (ESC) and
small-angle luminosity monitors (SAC)\null.  The AMY detector has been
described in detail elsewhere.\refmark{\rfAMY}

Final states with two or more photons are detected primarily by the
SHC\null.
This is a 14.4 radiation length gas-sampling calorimeter, consisting of
twenty alternating layers of lead and resistive plastic proportional
tubes.\refmark{\rfSHC}  Each tube layer is sandwiched between
cathode strips in the $\phi$ and $\theta$ directions.
The SHC is constructed in six sextants, each covering
an azimuthal range of $60^{\circ}$ and a polar angle range of
$|\cos\theta| < 0.73$.

Measurement of electromagnetic shower locations and energies in the SHC is
determined by the signals from the cathode strips.  The strips are arranged
in projective towers that subtend an angle of 14.2 mr from the interaction
point, and are ganged internally into five longitudinal divisions.
The empirical energy resolution is found to be $\Delta E/E = 0.23 / \sqrt{E} +
0.06$, with $E$ in GeV\null.  The large constant term is thought to be
due to the
effects of operating in the 3 T magnetic field.  Shower centroids are typically
located in $\phi$ and $\theta$ to less than one strip width.

\noindent\underbar{Triggering}

Within a given layer of the SHC, anode signals are ganged into 48 azimuthal
sections; the ganged signals are summed externally over multiple layers to
form trigger signals for 48 towers with either no segmentation or four
longitudinal divisions.  Three redundant combinations of these
signals provide triggering for $\ee \to \twog$ events:  (1) total
energy trigger---the analog sum of all 48 trigger signals exceeding 8 GeV,
(2) two-cluster trigger---two separate towers in coincidence with energies
exceeding 5 GeV for one and 3 GeV for the other, and (3) shower shape
trigger---coincidence between any two successive longitudinal divisions in
a given tower, or in neighboring towers, with total energy exceeding
3 GeV\null.


Bhabha electrons are used to check the triggering efficiency independently,
since they behave similarly to $\twog$ events in the SHC\null.  Bhabha events
are triggered
independently by the SHC triggers and by charged-track triggers derived from
signals in the CDC and ITC\null.  By examining the frequency of Bhabha
events with a track trigger but no SHC trigger, we found the SHC trigger
efficiency for Bhabha events to be essentially 100\%.  We assume,
therefore, that the SHC triggering efficiency for $\twog$ events exceeded
99\% for all of the data of this study.

\noindent\underbar{Luminosity} \underbar{Determination}

To search for an anomaly in the production of $\twog$ events, the most critical
parameter aside from event statistics is the reliable determination of the
integrated luminosity.  This is accomplished in the AMY experiment by recording
small-angle Bhabha events in the
ESC, a sampling calorimeter constructed of alternating layers of lead and
scintillator, with proportional tubes at the depth of electromagnetic shower
maximum for position determination.  The ESC covers the polar angle range
$0.799 < |\cos\theta| < 0.982$.
The systematic error in ESC luminosity is estimated to be $\sim 2$\%, and is
dominated by the precision on alignment and fiducial definition.

\chapter{\bf Event Selection and Monte Carlo}

The selection of $\ee \to n\gamma$ events $(n \ge 2)$ was
performed according to the following criteria:
(1) at least two SHC clusters with energies greater than $E_{beam}/3$ each,
(2) the polar angle for each such cluster in the range $45^{\circ} < \theta <
135^{\circ}$,
(3) the acollinearity angle between the two most energetic clusters of
at most $10^{\circ}$, and
(4) no charged tracks in the event.
The resulting sample of 1054 events was visually scanned, resulting in the
rejection of 19 background events from cosmic
rays, misidentified Bhabhas, and SHC noise.

The cross section for $\ee \to \twog$,
including radiative corrections to ${\cal O}(\alpha^3)$,
was calculated by Monte Carlo integration using a program by
Fujimoto, Igarashi, and Shimizu.\refmark{\rfFUIG,\rfFUIGSH}
This program is an implementation of the
BASES/SPRING Monte Carlo package by Kawabata.\refmark{\rfSKAW}
The cross section of $(34.81 \pm 0.18)\,\rm pb$ is
based on a fully simulated sample of $1.06\,\rm fb^{-1}$ at 58 GeV that was
subjected to the same selection criteria as the experimental data. This
value includes an event selection efficiency of
$\varepsilon_{\twog} = 0.933 \pm 0.011$.

We estimate the error in normalization for this analysis to be 2.3\%, taking
into account the uncertainty in the luminosity measurement and including
the error in the $\twog$ event selection efficiency.

\chapter{\bf Results}
Table~1 summarizes the results of the energy scan. The center-of-mass energies
$\sqrt{s}$
represent their actual values at the AMY interaction point rather than the
nominal accelerator energies. The quantity $\sigmaE$ represents the rms width
of the center-of-mass energy distribution.
The luminosities $L$ are those obtained with the ESC\null.
The quantity $N_{\twog}$ [$N_{QED}$] represents the number of observed
[simulated] $n\gamma$ $(n \ge 2)$ events that pass the event selection
criteria; $N_{QED}$ is
derived from the Monte Carlo simulation at 58 GeV discussed earlier, scaled by
$s$ and adjusted to the tabulated luminosity.
The ratio of the observed $\twog$ events to that expected from the Standard
Model is also shown for each center-of-mass energy.

Barrel Bhabha events obtained concurrently with $\twog$ events are used to
test the internal consistency of the runs and to provide an additional
check for systematic errors. For each center-of-mass energy, the ratio of
observed barrel Bhabha events to ESC luminosity is scaled by $s$.  These
quantities are given in the last column of Table~1, and indicate the
constancy in the operation of the SHC during the scan runs.

\chapter{\bf Analysis}

The ratio of observed $n\gamma$ events $(n \ge 2)$ to the QED expectation
is plotted as a function of center-of-mass
energy in Figure~1.  The vertical error bars on the ratios are statistical
only; the horizontal error bars indicate the rms spread in $\sqrt{s}$ rather
than the error on the central values.  A comparison of the data to QED alone,
allowing the normalization to float without constraint, gives a fully
consistent fit with
$\chi^2 / {\rm dof} = 0.68$ for 8 degrees of freedom and a normalization
value of $0.961 \pm 0.029$.  This fit is shown as the solid curve in Figure~1.
The deviation of the normalization from unity is consistent, given the 2.9\%
statistical uncertainty from the fit on this parameter as well as the 2.3\%
normalization uncertainty mentioned above.

To examine the effect of a new state $X$, we again allow the
normalization of the $\twog$ data to float without constraint, and then
compare these data to a
model in which the $\twog$ events are produced
by an $s$-channel resonance $X$ of mass $M$ and total width $\Gamma$ that
sums incoherently with the conventional QED processes.

The effective cross section for production of $X$ at center-of-mass energy
$\sqrt{s}$,
integrated over the solid angle of the SHC, is parameterized by a Breit-Wigner
form:
$$\sigmaX(s) = (2J+1) \Gee \BRgg f(s,M,\Gamma)\,,
\eqno(1)$$
where
$$f(s,M,\Gamma) = (\Omega\varepsilon)_{\rm eff}\,
                  {\pi \over s}\,
                  {\Gamma \over (M - \sqrt{s})^2 + \Gamma^2 / 4}\,\cdot
\eqno(2)$$
$\Gee$ is the partial width for $X \to \ee$, $\BRgg \equiv
\Ggg / \Gamma$ is the branching ratio for $X \to \twog$, and
$(\Omega\varepsilon)_{\rm eff}$ is the effective acceptance for detecting
$\ee \to \twog$.  We use $(\Omega\varepsilon)_{\rm eff} =
\Omega_{\rm SHC}\cdot\varepsilon_{\twog} = 0.707 \cdot 0.933 =
0.659$.  (This must be modified somewhat if the differential
cross section is not isotropic.)  The cross section $\sigmaX$ is then convolved
with the radiatively corrected beam resolution function\refmark{\rfJACKSON}
$$G_R(s,E) = \left( {2\sigmaE \over \sqrt{s} } \right)^t
             {t \over \sqrt{2\pi}\sigmaE}
             \int_0^{\,\,\infty} x^{t-1}\,\exp\left[ -(z-x)^2/2 \right]\,dx\,,
\eqno(3)$$
where $z = (\sqrt{s} - E) / \sigmaE$,\ \ $t = 2(\alpha/\pi)[\ln(s/m_e^2) -1]$,
and $\sigmaE$ is the width of the center-of-mass energy distribution from
Table~1, to give the observed effective cross section
$$\tilde\sigmaX(s) = (2J + 1)\Gee\BRgg
                      \int_{0}^{\,\,\infty}
f(E^2,M,\Gamma)\,G_R(s,E)\,dE\,\cdot
\eqno(4)$$

The number of events expected from this process
at center-of-mass energy $\sqrt{s}$ for an integrated luminosity $L$
is given by $N_X(s) = \tilde\sigmaX(s) \cdot L$, and adds to the number
expected from QED alone, $N_{QED}(s)$, listed in Table~1.

For a normalization $A_{\circ}$, the likelihood of observing
$N_X(s_i)$ events from $s$-channel production of $X$
at center-of-mass energy $\sqrt{s_i}$ is a rescaled Poisson
distribution:\refmark{\rfHELENE}
$$\eqalign{{\cal L}(s_i) &= A_{\circ}\,
   \exp\left\{ N_{QED}(s_i) - A_{\circ}[N_X(s_i) + N_{QED}(s_i)]\right\}\cr
&\qquad {}\times \left\{{A_{\circ}[N_X(s_i) +N_{QED}(s_i)] \over N_{QED}(s_i)}
                             \right\}^{\textstyle N_{\twog}(s_i)}\cr}
\eqno(5)$$
and the overall likelihood is ${\cal L} = \prod_{i} {\cal L}(s_i)\,.$

This likelihood function peaks at $M = (58.35^{+0.24}_{-0.08})\GeV$,
$\Gamma = (16\pm 13)\MeV$ and
$(2J+1)\Gee\BRgg = (4.2^{+3.4}_{-3.1})\keV$, with a normalization of
$A_{\circ} = 0.955 \pm 0.040$.
The best-fit curve of $A_{\circ}(N_X + N_{QED}) / N_{QED}$ is shown as the
dashed curve in Figure~1.
A comparison of this curve to the data has $\chi^2 / {dof} = 0.47$
with 5 degrees of freedom. Compared to $\chi^2 / {dof} = 0.68$ with 8
degrees of freedom for QED alone, we conclude that this peak is not
statistically significant.

We extract 90\% confidence level upper limits on $(2J+1)\Gee\BRgg$, for
the normalization of 0.955 and particular choices of $M$ and $\Gamma$,
by numerical integration of the likelihood function.
\footnote{\dag} {Since the data are consistent with pure QED, the
confidence {\it interval} includes the point \hbox{$(2J+1)\Gee\BRgg=0$.}}
These limits are shown in Figure~2.  It is apparent
that the limits are insensitive to the total width $\Gamma$ when it is
comparable to or smaller than the center-of-mass energy spread $\sigmaE$\null.
Furthermore, the limits shift only slightly if the normalization is varied
from its maximum likelihood value.


Upper limits on $(2J+1)\Gee\BRgg$ at the 90\% confidence level range from
0.5 keV to 8 keV, for an $X$ mass between 57 and 59.6 GeV and a total width
$\Gamma$ below 100 MeV\null.  Our limits can
be compared directly with the expectation from the L3 observation of
$\ee \to \ll\twog$ under the assumption that their clustered events arose from
the production of a new state $X$\null.  For example,  if we assume
that $X$ is a
scalar that couples only to photon and lepton pairs with a total
width of $\Gamma \equiv \Ggg + 3\Gee$, then an
upper limit on $(2J+1)\Gee\BRgg$
can be converted into an exclusion region in the plane of
$\Ggg$ vs.\ $\Gee$. Our 90\% C.L. upper limit is
shown in Figure~3
for a scalar $X$ of mass 59 GeV\null.

Our limits may be incorporated with 90\% C.L. limits of other
experiments such as the L3 measurement\refmark{\rfLTHREE} of
$\ee\to\ll\twog$ and the OPAL measurement\refmark{\rfOPAL} of
$\ee\to\threeg$. For the former case, we calculate\refmark{\rfKURIHARA}
the cross section
using the automatic amplitude generator GRACE\refmark{\rfGRACE}
including all possible diagrams. The limits from the three events with
$M_{\twog}$ clustered near 59 GeV observed by the L3 collaboration are
shown as the dashed curve in Figure~3, while the limits imposed by the
absence of a $\ll$ invariant mass near 59 GeV are given by the
dash-dotted curve. The limits set by the non-observation of an excess
of events with $M_{\twog}=59$ GeV in the OPAL $\threeg$ measurement are
shown as the dotted curve of Figure~3. The limits of our measurement are
thereby restricted to satisfying the condition that
$\Gamma_{\twog}<10.6$ MeV and $\Gamma_{ee}<120$ MeV\null. To summarize, the
four {\it inclusion} regions overlap in the lower left of Figure~3.

For a (pseudo-)scalar that couples directly to pairs of fermions as well
as photons, the exact forms of the decay widths for $X \to \ee$ and
$X \to \twog$ are
$$\Gee = {g_{\rm ee}^{2}\over {8\pi}} M
\eqno(6)$$
and
$$\Ggg = {\alpha^{2}\over{64\pi^{3}}} {M^{3}\over F_{X}^2}
\eqno(7)$$
respectively, where $g_{\rm ee}$ is the coupling constant of $X$ to electrons,
$\alpha$ is the fine structure constant, and $F_X$ is a mass parameter
analogous to $f_{\pi}$ in $\pi^{\circ}$ decay. We use
$\alpha / (\pi F_X)$ as the coupling constant of $X$ to $\twog$.
Assuming that $\BRgg \sim 1$, we can convert the 90\% C.L. upper limit on
$\Gee\BRgg$ from our likelihood analysis into an upper limit on $g_{\rm ee}$
of $1.78\times 10^{-3}$.  We also place a 90\% C.L. lower limit
on $F_X$ of 730 MeV that comes from the OPAL upper limit on $\Gamma_{\twog}$
of 10.6 MeV.\refmark{\rfOPAL}
We examine the effect of these limits upon the $ee\gamma$
vertex correction value, $(g - 2)_e$, which will have an additional
contribution
from this model of the form\refmark{\rfAGUIL}
$${\cal A}_X \simeq {m_e\over F_X}{\alpha g_{\rm ee}\over 4\pi^3}
\ln {\Lambda\over M}
\eqno(8)$$
where $\Lambda$ is the ultraviolet cutoff (fixed at 1 TeV)\null.  Our limits
on $F_X$ and $g_{\rm ee}$ imply an upper bound of
${\cal A}_X < 2.0 \times 10^{-10}$,
which is within the maximum allowed
contribution\refmark{\rfKINOSH} to $(g - 2)_e$ of $2.7 \times 10^{-10}$.

\chapter{\bf Conclusion}

We have searched for resonant production of a new state $X$ in $\ee$
collisions that couples to photon pairs at center-of-mass energies between
57.4 and 59.5 GeV\null.  We find that the observed data are consistent with the
QED prediction for $\ee \to \twog$ ($\chi^2/{dof} = 0.68$ for 8 degrees of
freedom). Furthermore, we have extracted
90\% C.L. upper limits on $(2J+1)\Gee\BRgg$ of 0.5---$8.0\keV$
for the process $\ee \to X \to \twog$, should a state $X$ exist with a mass
between 57 and $59.6\GeV$\null.

\chapter{\bf Acknowledgement}

We thank the TRISTAN staff for the excellent operation of the storage
ring. In addition we acknowledge the strong support provided by the
staffs of our home institutions.  We are particularly indebted to
Professor Lay Nam Chang for fruitful discussions on the theoretical
aspects of this paper.  This work has been supported by the Japan
Ministry of Education, Science and Culture (Monbusho), the Japan Society
for the Promotion of Science, the U.S.\ Department of Energy, the U.S.
National Science Foundation, the Korean Science and Engineering
Foundation, the Ministry of Education of Korea and the Academia Sinica
of the People's Republic of China.
\vfill\endpage
\refout
\vfill\eject
\nopagenumbers
\centerline {{\bf TABLE CAPTIONS}}
\vskip 0.5 cm

\noindent{\bf Table 1.}  Results of the energy scan. $\sqrt{s}$ is the
center-of-mass energy, $\sigmaE$ is the width of the center-of-mass energy
distribution,
$L$ is the integrated luminosity from the ESC, $N_{QED}$ is the number
of events expected from QED alone, $N_{\twog}$ is the number of $\ee \to \twog$
seen, and $N_{BB}^{\ast}/L$ is the number of barrel Bhabha events seen per
$\rm pb^{-1}$ of luminosity, scaled by $s$.
\endpage
\vfill\eject
\noindent\kern-20truept\vbox{\tabskip=0pt \offinterlineskip
\def\tablerule{\noalign{\hrule}}
\setbox0=\hbox{$88.88$}           
\setbox1=\hbox{$888.8 \pm 8.8$}   
\setbox2=\hbox{$888$}             
\setbox3=\hbox{$260.9 \pm 18$}    
\setbox4=\hbox{18}                
\setbox5=\hbox{260.9}             
\def\LUM#1{\hbox to \wd0{\hfil $#1$}}
\def\NQED#1{\hbox to \wd1{\hfil $#1$}}
\def\NGG#1{\hbox to \wd2{\hfil $#1$}}
\def\NBB#1#2{\hbox to \wd3{\hfil $\hbox to \wd5{#1\hfil} \pm
\hbox to \wd4{\hfil #2}$}}
\halign{%
\vrule\enskip\hfil$\mathstrut #$\hfil\enskip&%
\vrule\enskip\hfil$#$\hfil\enskip&%
\vrule\enskip\hfil$#$\hfil\enskip&%
\vrule\enskip\hfil$#$\hfil\enskip&%
\vrule\enskip\hfil$#$\hfil\enskip&%
\vrule\enskip\hfil$#$\hfil\enskip&%
\vrule\enskip\hfil$#$\hfil\enskip\vrule\cr
\tablerule
\omit\vrule height 4pt&&&&&&\cr
\sqrt{s}\ \rm (GeV) & \sigmaE\ \rm(GeV) & L\ (\rm pb^{-1}) &
N_{QED} & N_{\twog} & N_{\twog}/N_{QED} & N_{BB}^{\ast}/L \cr
\omit\vrule height 4pt&&&&&&\cr
\tablerule
\omit\vrule height 4pt&&&&&&\cr
57.374 & 0.095 &  \LUM{2.05} &  \NQED{72.9 \pm 0.4} &  \NGG{69} &
                                0.95 \pm 0.11  &   \NBB{265}{11}  \cr
\omit\vrule height 4pt&&&&&&\cr
57.772 & 0.096 & \LUM{19.55} & \NQED{680.5 \pm 3.4} & \NGG{636} &
                               0.935 \pm 0.037 &
                               \NBB{276.0}{\rlap{3.8}\hfill} \cr
\omit\vrule height 4pt&&&&&&\cr
57.972 & 0.096 &  \LUM{1.39} &  \NQED{48.4 \pm 0.2} &  \NGG{49} &
                                1.01 \pm 0.15  &   \NBB{272}{14}  \cr
\omit\vrule height 4pt&&&&&&\cr
58.220 & 0.098 &  \LUM{1.33} &  \NQED{45.8 \pm 0.2} &  \NGG{50} &
                                1.09 \pm 0.15  &   \NBB{259}{14}  \cr
\omit\vrule height 4pt&&&&&&\cr
58.470 & 0.098 &  \LUM{1.33} &  \NQED{45.6 \pm 0.2} &  \NGG{54} &
                                1.18 \pm 0.16  &   \NBB{308}{15}  \cr
\omit\vrule height 4pt&&&&&&\cr
58.718 & 0.099 &  \LUM{1.73} &  \NQED{58.7 \pm 0.3} &  \NGG{61} &
                                1.04 \pm 0.13  &   \NBB{278}{13}  \cr
\omit\vrule height 4pt&&&&&&\cr
58.968 & 0.100 &  \LUM{1.36} &  \NQED{45.9 \pm 0.2} &  \NGG{45} &
                                0.98 \pm 0.15  &   \NBB{284}{15}  \cr
\omit\vrule height 4pt&&&&&&\cr
59.216 & 0.100 &  \LUM{1.21} &  \NQED{40.5 \pm 0.2} &  \NGG{45} &
                                1.11 \pm 0.17  &   \NBB{294}{16}  \cr
\omit\vrule height 4pt&&&&&&\cr
59.466 & 0.102 &  \LUM{0.97} &  \NQED{32.3 \pm 0.2} &  \NGG{26} &
                                0.81 \pm 0.16  &   \NBB{236}{16}  \cr
\omit\vrule height 4pt&&&&&&\cr
\tablerule}  
}  
\vfill\eject
\nopagenumbers
\Twelvepoint
\centerline {{\bf FIGURE CAPTIONS}}
\vskip 0.5 cm

\noindent{\bf Figure 1.}  Ratio of observed $\twog$ events to the QED
prediction.  The solid curve is the
QED prediction, with $\chi^2 / dof = 0.68$ for 8 degrees of freedom
and a normalization of 0.961.  The dashed curve is the maximum likelihood
prediction, with $\chi^2 / {dof} = 0.47$ for 5 degrees of freedom and a
normalization of 0.955.

\noindent{\bf Figure 2.} 90\% C.L. upper limits on $(2J+1)\Gee\BRgg$
as a function of mass $M$ of the state $X$, for widths of 1 MeV (solid curve),
100 MeV (dashed curve), 1 GeV (dotted curve) and 3 GeV (dash-dotted curve),
with a normalization of 0.955.  The region above a given curve is excluded.

\noindent{\bf Figure 3.} Exclusion regions (90\% C.L.) on $\Gamma_{\twog}$
{\it vs.} $\Gamma_{\rm ee}$ for a scalar $X$ with $M \sim 59\GeV$,
assuming $\Gamma \equiv \Gamma_{\twog} + 3\Gamma_{\rm ee}$.  The region above
and to the right of a given curve is excluded.  The solid curve is from this
experiment.  The other curves are from the observation of up to three
$\ll\twog$ events with $M_{\twog} \sim 59\GeV$ (dashed) and the absence
of $\ll\twog$ events with $M_{\ll} \sim 59\GeV$ (dash-dotted) (Ref.~\rfLTHREE),
and from the lack of a signal at $M_{\twog} \sim 59\GeV$ in $\threeg$ events
(dotted) (Ref.~\rfOPAL).

\endpage
\end